%% file: main.tex
\documentclass[runningheads]{llncs}
\usepackage{appendix}
\usepackage[table]{xcolor}% http://ctan.org/pkg/xcolor
\usepackage{graphicx}
\usepackage{subcaption} % For sub-figures
\usepackage{amsmath}

\usepackage{comment}
\usepackage{multirow}
\usepackage[hidelinks]{hyperref}
\usepackage{url}
\usepackage[export]{adjustbox}
\usepackage{array, caption, tabularx,  ragged2e,  booktabs}
\usepackage{color,soul}
\usepackage[inline,shortlabels]{enumitem}
\usepackage{multicol}
\usepackage{tabularx}

\usepackage[normalem]{ulem}

\usepackage{ntheorem}
\theoremseparator{:}
\newtheorem{hyp}{Hypothesis}

\newcommand{\para}[1]{\textbf{#1}.}

\definecolor[named]{mygreen}{rgb}{0.4,0.9,0.7}
\definecolor[named]{myred}{rgb}{1,0.7,0.5}

\let\oldpara\para
\renewcommand{\para}{\smallskip\noindent\oldpara}

\begin{document}
%
% \title{Contribution Title\thanks{Supported by organization x.}}
\title{Combining Multiple Query Performance Prediction Methods for Enhanced Accuracy}
\title{Predictor Fusion Revisited: Enhancing Accuracy through Method Combination}
\title{Does Combining Multiple Query Performance Predictors Improve Accuracy?}
% just a placeholder
\title{Combining Query Performance Predictors:\\ A Reproducibility Study}
\titlerunning{Combining Query Performance Predictors: A Reproducibility Study}

% \author{}
\author{
Sourav Saha
\inst{1}
\orcidID{0000-0001-8091-0685}
Suchana Datta
\inst{2}
\orcidID{0000-0001-9220-6652}
\and
Dwaipayan Roy
\inst{3}
\orcidID{0000-0002-5962-5983} \and
Mandar Mitra
\inst{1}
\orcidID{0000-0001-9045-9971} \and
Derek Greene
\inst{2}
\orcidID{0000-0001-8065-5418}
}
\authorrunning{S. Saha et al.}
\institute{
	Indian Statistical Institute, Kolkata, India \and
	University College Dublin, Ireland \and
	Indian Institute of Science Education and Research, Kolkata, India\\
	\email{souravsaha.juit@gmail.com},
	\email{suchana.datta@ucd.ie},
	\email{dwaipayan.roy@iiserkol.ac.in},
	\email{mandar@isical.ac.in},
	\email{derek.greene@ucd.ie}
}

\maketitle              % typeset the header of the contribution
\setcounter{footnote}{0}

\input{ecir25/abstract}
%\input{intro.tex}
\input{ecir25/intro-alt.tex}
\input{ecir25/relwork}

\input{ecir25/results}
\input{ecir25/analysis}

\input{ecir25/analysis2-mm}
\input{ecir25/conclusion}

\subsubsection*{\textbf{Acknowledgement.}}
\small This publication is partially funded by (i) the European Research Council (ERC) under the Horizon 2020
research and innovation program (Grant No. 884951); (ii) Science Foundation Ireland (SFI) to the Insight Centre for Data Analytics under grant No 12/RC/2289\_P2. Sourav Saha acknowledges the support from the TCS Research Scholar program.

% \newpage

\bibliographystyle{splncs04}
% \typeout{}
\bibliography{refs}
\end{document}

%% file: ecir25/abstract.tex
\begin{abstract}
  A large number of approaches to Query Performance Prediction (QPP) have
  been proposed over the last two decades. As early as 2009, Hauff et al.~\cite{hauff2009combination} explored whether different QPP methods
  may be combined to improve prediction quality. Since then, significant
  research has been done both on QPP approaches, as well as their
  evaluation. This study revisits Hauff et al.'s work to assess the
  reproducibility of their findings in the light of new prediction methods,
  evaluation metrics, and datasets. We expand the scope of the earlier
  investigation by: (i)~considering post-retrieval methods, including
  supervised neural techniques (only pre-retrieval techniques were
  studied in~\cite{hauff2009combination}); (ii)~using \emph{sMARE} for
  evaluation, in addition to the traditional correlation coefficients and
  \emph{RMSE}; and (iii)~experimenting with additional datasets (Clueweb09B and
  TREC DL). Our results largely support previous claims, but we also
  present several interesting findings. %\inote{\textcolor{red}{eta baad to?}Specifically, two
    %pre-retrieval approaches, \emph{MaxIDF} and \emph{AvgIDF}, outperform
    %unsupervised post-retrieval predictors, and perform at par with neural
    %approaches in terms of RMSE. Also, combining post-retrieval approaches
    %significantly improves RMSE on the Robust collection.} 
    We interpret these findings by taking a more nuanced look at the correlation
  \emph{between} QPP methods, examining whether they capture diverse
  information or rely on overlapping factors.

\keywords{Query Performance Prediction \and Fusion \and Evaluation \and Reproducibility \and Empirical Study.}
\end{abstract}

%%% Local Variables:
%%% mode: LaTeX
%%% TeX-master: "main"
%%% End:

%% file: ecir25/intro-alt.tex
\section{Introduction}\label{sec:intro}

Query Performance Prediction (QPP) has long been an important area of study within information retrieval (IR)~\cite{carmel_sigir10,townsend_2002_clarity,precision_prediction,survey_preret_qpp,kurland_unified_2011,qpp_ref_list_sigir17,uef_kurland_sigir10,kurland_tois12,zhou:2006,zhou_2007_qpp}. QPP focuses on estimating (without utilizing relevance judgments) how well a search engine will perform on a given query. 
Accurate predictions of query performance can guide decisions regarding search strategies, resource allocation, and query reformulation~\cite{Feild,Li:2012,RhaSB17,querysuggestion}. 
Over the years, various QPP approaches have been proposed \cite{townsend_2002_clarity,qpp_preret_ecir08,banerjee03extendedgloss,rada:avpath,patwardhan:avvp,qpp_coherence,mothe_linguistics,amati_2004_difficulty,townsend_2004_drift,carmel_2006_what,zhou_2007_qpp,shtok_qpp_ref_list,qpp_ref_list_sigir17,kurland_tois12}. These methods can be broadly categorized as pre-retrieval or post-retrieval techniques. 
Pre-retrieval methods~\cite{survey_preret_qpp,qpp_preret_springer,qpp_preret_ecir08} estimate query difficulty using only the query and collection statistics, while post-retrieval methods~\cite{townsend_2002_clarity,kurland_tois12,zhou_2007_qpp,uef_kurland_sigir10,roitman_scnqc,qpp_ref_list_sigir17,shtok_qpp_ref_list} also use retrieval results (ranked lists of documents) to make predictions.

Given the plethora of methods proposed in the literature, researchers have also investigated whether combining different QPP methods yields better predictions~\cite{vinay:2006,zhou_2007_qpp,hauff2009combination,uef_kurland_sigir10,kurland_unified_2011,kurland_back_2012}.
In a comprehensive study involving 19 pre-retrieval predictors (but no post-retrieval methods) and 3 test collections, {Hauff et al.}~\cite{hauff2009combination} showed that combining predictors yields significant improvements in correlation between predicted and actual performance. 
However, the observed gains in terms of Root Mean Squared Error (\emph{RMSE}) were modest, or even non-existent. 
Based on these observations, the authors proposed two important guidelines regarding QPP evaluation:
\begin{enumerate}[~i)]
% \begin{enumerate}[leftmargin=*]
\item When reporting the correlation\footnote{\scriptsize In Sections~\ref{sec:intro}--\ref{sec:rel-work}, we abuse notation, and use $\rho$ as an abbreviation for any of the correlation coefficients (Pearson's or Spearman's Rho and Kendall's Tau) that are commonly used when evaluating QPP methods. 
In Sections~\ref{sec:exp}--\ref{sec:discussion} (Experimental setup, results and analysis), $\rho$ and $\tau$ are used in accordance with standard convention.} between predictions and actual performance, a 95\% confidence interval for the observed correlation should also be reported.
This helps in assessing whether observed differences between methods are statistically significant.
\item In addition to $\rho$, a listwise metric, the Squared Error (SE) between the predicted and measured retrieval effectiveness should be used as a more interpretable, pointwise metric for QPP, with Root Mean Squared Error (\emph{RMSE}) representing the aggregation of these values across a test collection.
\end{enumerate}

In the 15 years since the study by {Hauff et al.}, QPP has witnessed a
great deal of research activity. Our study revisits QPP method fusion in
the light of these advancements, in order to address the following research
questions:
\begin{itemize}
  \setlength{\leftmargin}{2cm} % Adjust the value as needed
\item[\textbf{RQ1 - }] Does the combination of pre-retrieval QPP methods
  consistently enhance correlation with Average Precision (\emph{AP})
  across different datasets, including recent web collections, as suggested
  in~\cite{hauff2009combination}? Additionally, does combining
  predictors significantly affect \emph{RMSE}? (Hauff et
  al.'s results suggest a negative answer to this question.)
  % , or does \emph{RMSE} similarly vary across
  % newer datasets?
\item[\textbf{RQ2 - }]
% Do post-retrieval approaches exhibit similar
%   patterns? Specifically, 
% Does combining multiple post-retrieval QPP
  % estimators yield a fused predictor that outperforms the individual
  % post-retrieval approach across different datasets? 
  % Furthermore, do the performance trends of these
  % two types of predictors differ across datasets, and can we account for
  % any observed differences?
Do the answers to \textbf{RQ1} change if we consider post-retrieval predictors instead of pre-retrieval estimators?
\item[\textbf{RQ3 - }] 
% Following from the above, 
Is there a relationship between the predictions of different QPP methods, both within and
  across pre-retrieval and post-retrieval categories?
\end{itemize}
To answer the above questions, 
and to assess whether Hauff et al.'s
observations generalize to newer QPP methods, evaluation metrics, and
more modern datasets, 
we expand the scope of our experiments along
the directions listed below:
\begin{itemize}
\item \textbf{Incorporation of post-retrieval methods:} While {Hauff et al.}~\cite{hauff2009combination}
  limited their analysis to pre-retrieval methods, we also consider
  post-retrieval methods. This includes not only traditional unsupervised methods but also more recent supervised neural approaches.
  % providing a broader picture of the effectiveness of QPPs together with
  % their combinations.
\item \textbf{Evaluation metrics:} When measuring prediction quality, we use
  \emph{scaled Mean Absolute Rank Error}
  (\emph{sMARE})~\cite{qpp_new_eval_ecir21}, in addition to the conventional
  correlation measures and \emph{RMSE}. \emph{sMARE} is a pointwise metric that is seeing
  increased use in recent QPP research. One advantage of \emph{sMARE} over \emph{RMSE} is
  that calculating \emph{RMSE} involves fitting a regression model, since QPP
  scores are generally not intended to be interpreted as predicted values
  of standard metrics like Average Precision (AP), and are often not even in the $[0,1]$ range;
  in contrast, \emph{sMARE} is parameter free.
\item \textbf{Test collections:} {Hauff et al.} used one news collection
  (TREC 678) and two web collections (WT10G and Gov2) for their experiments.
  In our study, we repeat their experiments on the TREC 678 collection
  (also reporting figures on the Robust collection with 100 more queries),
  and extend the evaluation to ClueWeb09B, a much larger Web collection, as
  well as the more modern MS MARCO collections, thus providing a more comprehensive evaluation.
\end{itemize}
Thus, our study has elements of both reproducibility (different team, same
experimental setup) and replicability (different team, different
experimental
setup)\footnote{\scriptsize\url{https://www.acm.org/publications/policies/artifact-review-and-badging-current}}.
Given that the original experiments that we seek to reproduce and extend
are 15 years old, we have not attempted to compute quantitative reproducibility metrics
for our
efforts using recently proposed
tools~\cite{breuer2020measure,maistro2023depth}. Instead, when summarizing the answers to our research questions later in Section~\ref{conclusion}, we categorize our observations as either new insights, confirmations of, or deviations from those presented in~\cite{hauff2009combination}.
A GitHub
repository\footnote{\scriptsize\url{https://github.com/souravsaha/qpp-comb}}
provides the artifacts (retrieval results/ ranked lists, code for QPP
methods, fusion, and subsequent analysis) necessary for other researchers
to independently corroborate our findings.

%%% Local Variables:
%%% mode: LaTeX
%%% TeX-master: "main"
%%% End:

%% file: ecir25/relwork.tex
\section{Related Work}
\label{sec:rel-work}

\subsection{An overview of QPP methods} \label{ss:qpp}
% As mentioned in Section~\ref{sec:intro}, QPP approaches are generally
% categorized as either \emph{pre-retrieval} or \emph{post-retrieval} methods. 
In this section, we provide a brief review of two categories of QPP methods, namely, \emph{pre-retrieval} and \emph{post-retrieval} QPPs. 

\para{Pre-retrieval approaches} Pre-retrieval performance predictors can,
in turn, be classified based on the query term features used to compute
them. One category of predictors estimates query-term \emph{specificity}:
they assume that more discriminative terms (i.e., those with higher inverse
document frequency (IDF) or inverse collection frequency (ICTF)) are
better at identifying relevant content.
Average IDF (\texttt{avgIDF}) and the maximum
IDF values (\texttt{MaxIDF})~\cite{townsend_2002_clarity} fall under this category. 
Some studies compute the standard deviation of IDF from the average of ICTF
values (\texttt{AvgICTF}) as specificity-based predictors~\cite{qpp_coherence}. Zhao
et al.~\cite{qpp_preret_ecir08} proposed predictors using collection-based
specificity to predict query difficulty, including methods like the sum of
collection and query term similarity (\texttt{SumSCQ}), normalized query and
collection similarity (\texttt{AvgSCQ}), and maximum collection and query term
similarity (\texttt{MaxSCQ}).

Another category is \emph{rank sensitivity}, which estimates query difficulty based on the variation of term weights across documents. Common techniques here include
% The premise is that an even distribution of query term weights makes it hard to distinguish relevant from non-relevant documents~\cite{qpp_preret_ecir08}.
% Based on sensitivity, the researchers have introduced methods 
\texttt{SumVAR} (the total of query term weight variations), \texttt{AvgVAR} (the normalized sum of these variations), and \texttt{MaxVAR} (the maximum variation among the terms)~\cite{qpp_preret_ecir08}.
\emph{Term relatedness} predictors leverage estimates of how related the query terms are to each other.
Methods like \texttt{AvLesk}~\cite{banerjee03extendedgloss}, \texttt{AvPath}~\cite{rada:avpath}, and \texttt{AvVP}~\cite{patwardhan:avvp} use external resources, typically WordNet~\cite{wordnet} or co-occurrence statistics, to determine semantic distances between query terms. 
The presence of an ambiguous term in a query can lead to imprecise information needs~\cite{sanderson1994word},
% , as an ambiguous term may have multiple relevant senses~\cite{sanderson1994word}. 
often resulting in difficulty for retrieval functions~\cite{carmel_sigir10,mothe_linguistics}.
Predictors in this category measure \emph{ambiguity} by computing the average inter-document similarity of all pairs of documents containing at least one query term~\cite{qpp_coherence} at a cost of quadratic time complexity which can be mitigated by sampling a subset of document pairs~\cite{hauff2009combination}.
Variants like \texttt{AvQCG} include pairs containing all query terms. 
% However, a significant drawback of this inter-document method is its quadratic time complexity, which can be mitigated by sampling a subset of document pairs~\cite{hauff2009combination}.
% Some existing approaches further utilize WordNet to assess query ambiguity, such as 
Averaged Polysemy (\texttt{AvP}) and Averaged Noun Polysemy (\texttt{AvNP}) utilize WordNet to assess query ambiguity by considering the number of senses associated with query terms~\cite{mothe_linguistics}.
% For our empirical study in this paper, we make use of different QPP techniques chosen from each class (more details in Section~\ref{sec:exp}).
% A performance comparison of these pre-retrieval predictors can be found in~\cite{Hauff:2008}, with further details available in~\cite{carmel:monograph,hauff:thesis}.

\para{Post-retrieval approaches}
The post-retrieval predictors infer their predictions by taking the ranked
list of documents in response to the query as input. These can be broadly
% Depending on which aspects are considered to compute the prediction, there are 
classified into \emph{coherence}-based, \emph{score}-based, and
\emph{robustness}-based predictors. Coherence-based predictors measure how
strongly documents retrieved are clustered together, such as
\texttt{Clarity}~\cite{townsend_2002_clarity}.
This pioneering approach influenced a series of studies that built on the idea of query clarity~\cite{amati_2004_difficulty,carmel_2006_what,townsend_2004_drift,zhou_2007_qpp}.
{Score-based} predictors, e.g., \texttt{WIG}~\cite{zhou_2007_qpp}, \texttt{NQC}~\cite{kurland_tois12}, and \texttt{SMV}~\cite{qpp_sd_cikm14}, employ heuristics related to the retrieval scores of retrieved documents. 
Finally, {robustness-based} predictors, e.g., the \texttt{UEF}
family~\cite{uef_kurland_sigir10}, the Reference Lists framework~\cite{qpp_ref_list_sigir17,shtok_qpp_ref_list}, and \texttt{RSD}~\cite{roitman_ictir17}, compare the original ranking of documents with one produced by introducing noise in the query, the index, or documents.
% \inote{eita ki ekhane bola hobe?}
% For our study in this paper, we choose one or more QPP methods from each of the three classes.
In parallel, several researchers explored robustness-based predictors to assess result quality, focusing on the stability of the top-ranked documents~\cite{aslam:2007,kurland_qpp_rlm,vinay:2006,yom-tov_learning_2005,zhou:2006}.

% The foundational work in the field of query performance prediction emerged with~\cite{townsend_2002_clarity}, where \emph{query clarity} was estimated by calculating the Kullback-Leibler divergence between language models derived from search results and those from the entire corpus.

%
% Other QPP approaches, including the use of reference lists~\cite{shtok_qpp_ref_list,qpp_ref_list_sigir17} and query expansion~\cite{MM_qe_qpp,qpp_microblog_IPM}, have also been employed in several studies with positive outcomes.
%
% A notable advancement in the field came with score variance-based methods such as NQC~\cite{shtok_2009_nqc,kurland_tois12}.

% \textcolor{red}{ekhane ekta paragraph dhukbe : SD} 
More recently, researchers show that supervised QPP approaches outperform their unsupervised counterparts~\cite{bertqpp,group_bert_qpp22,deepqpp_datta22,qppbertpl_datta22,bertpe,neuralqpp}.
These methods, in general, use deep learning techniques trained on large-scale datasets, resulting in improved accuracy.
% However, they raise challenges regarding reproducibility due to their reliance on labeled data and complex architectures.
% In the supervised QPP paradigm, there are mainly two lines of research, (1) QPP with neural models, where the neural architectures are leveraged to estimate the prediction scores, such as \texttt{NeuralQPP}~\cite{neuralqpp}, \texttt{Deep-QPP}~\cite{deepqpp_datta22}, \texttt{qppBERT-PL}~\cite{qppbertpl_datta22}, \texttt{BERT-QPP}~\cite{bertqpp}, \texttt{GroupQPP}~\cite{group_bert_qpp22} etc., and (2) QPP for neural models, in which case QPP models are tailored for neural re-rankers, such as \texttt{WRIG}~\cite{datta23_tois_wrig}, \texttt{QPP-PRP}~\cite{asingh_qppprp_sig23}. This study  
% The introduction of 
The first (weakly) supervised approach, \texttt{NeuralQPP}~\cite{neuralqpp}, uses pairs of queries to learn a binary indicator denoting which one of the pair leads to a better (or worse) retrieval effectiveness.
% , has brought new capabilities to post-retrieval performance prediction. 
% A notable drawback of this pairwise technique is that given the training set size, the number of pairs is quadratic, leading to a significant increase in training time. 
A key drawback of this pairwise technique is that, depending on the size of the training set, the number of pairs grows quadratically, resulting in a substantial increase in training time.
A solution to this increased training time problem was proposed in~\cite{bertqpp} that introduces \texttt{BERT-QPP}, an adoption of contextual embeddings to perform pointwise QPP.
% Different from \texttt{NeuralQPP}, Datta et al. proposed to make use of interaction between query and document terms as signals for QPP. 
% More specifically, they employ 3-dimensional convolutional neural networks with shared parameters to train an end-to-end pairwise predictor, called \texttt{Deep-QPP}~\cite{deepqpp_datta22}. 
In contrast, the authors in~\cite{deepqpp_datta22} proposed \texttt{Deep-QPP}, a model employing a 3-dimensional convolutional neural network to train an end-to-end pairwise predictor.
% Inspired by the time efficiency of cross-encoder based \texttt{BERT-QPP}, Datta et al. later proposed \texttt{qppBERT-PL}~\cite{qppbertpl_datta22} that makes use of cross-encoding based query-document interactions in the form of BERT vectors, trained pointwise on individual queries but listwise by splitting the top documents into chunks.
Inspired by the time efficient \texttt{BERT-QPP}, researchers have further proposed \texttt{qppBERT-PL}~\cite{qppbertpl_datta22} which makes use of cross-encoding based query-document interactions in the form of BERT vectors, trained pointwise on individual queries while employing a listwise strategy by dividing the top documents into chunks.

\subsection{Combining QPP estimators}

While many QPP methods have been proposed, relatively less effort has been dedicated to effectively combining them.
Empirical studies have consistently shown that combining pre-retrieval and post-retrieval predictors can improve the accuracy of query performance prediction.
An early study proposed
% Yom-Tov et al. propose a model combining 
a model to enhance prediction by combining 
pre-retrieval features (e.g., the rounded logarithm of document frequency), with post-retrieval features, including overlap between ranked lists generated from sub-queries~\cite{yom-tov_learning_2005}. 
% This integration, achieved through histogram-based and decision tree-based frameworks, was shown to enhance predictive power~\cite{yom-tov_learning_2005}.
%
The first exploratory analysis of combined predictors was reported by Hauff
et al.~\cite{hauff2009combination}. Their experiments with
pre-retrieval predictors showed that fusing improves the performance of the
predictors in terms of correlation coefficients (denoted $\rho$).
% by Hauff et al. that the fusion techniques applied to QPP methods improve performance in terms of linear correlation coefficients
% 
% where Hauff et al. explored the potential benefits of combining QPP methods to improve prediction accuracy.
% Experimentation with pre-retrieval techniques led to the conclusion by Hauff et al. that the fusion techniques applied to QPP methods improve performance in terms of linear correlation coefficients.
%
% Further work by Kurland and colleagues explored integrating pre-retrieval and post-retrieval predictors within a probabilistic framework.
Subsequent research investigated the integration of pre-retrieval and post-retrieval predictors within a probabilistic framework.
The methods proposed in~\cite{kurland_unified_2011,kurland_back_2012}
achieved improved prediction accuracy by combining post-retrieval predictors. %, such as \texttt{Clarity Score}~\cite{townsend_2002_clarity} and Drift-based~\cite{kurland_tois12} predictors, highlighting the synergistic potential of these methods in query performance prediction.
% Their results indicated that the combined approach consistently outperformed the individual predictors~\cite{kurland_back_2012}.
Kurland et al. also demonstrated improvements in predictive accuracy through the combination of post-retrieval predictors, such as Clarity Score~\cite{townsend_2002_clarity} and Drift-based~\cite{townsend_2004_drift} predictors, highlighting the synergistic potential of these methods in QPP~\cite{kurland_unified_2011}.
Other research has approached QPP enhancement through the fusion of scores from multiple rankings derived from different retrieval models.
In~\cite{diaz_performance_2007}, the authors combined regularized document scores across multiple rankings, further substantiating the benefit of predictor fusion.
Additional studies support this trend, with researchers showing that combining post-retrieval predictors outperforms the use of individual methods, reinforcing the value of integrated predictor models in query performance prediction~\cite{raiber_setting,zhou_2007_qpp,hauff_dissertation,shtok_qpp_ref_list,roy_gmm-qpp}.
% In recent studies, researchers have integrated traditional pre-retrieval metrics with the Normalized Query Commitment (\texttt{NQC})~\cite{shtok_2009_nqc,kurland_tois12} which is a post-retrieval method, demonstrating a marked improvement over the individual predictors alone~\cite{roy_gmm-qpp}.

Our work builds upon these studies by extending the scope of predictor fusion, specifically focusing on post-retrieval methods, including modern neural approaches.
Additionally, we explore the reproducibility of earlier findings using larger collections, such as ClueWeb09B~\cite{web_track_10} and TREC DL~\cite{msmarco-data}.
By addressing gaps in the literature, 
%and applying rigorous reproducibility measures, 
we aim to provide a better understanding of the effectiveness of combined QPP methods. 

\subsection{Evaluation measures}
\label{ss:eval} 
Evaluating the effectiveness of QPP methods requires robust metrics that
capture various aspects of predictive accuracy and randomness. Over time,
several metrics have become standard in QPP research, each offering unique
insights into different facets of predictor performance. 
The primary metrics we apply for evaluation include Pearson's $\rho$ and
Kendall's $\tau$, 
\emph{RMSE}, %a squared error-based metrics,
together with the recently proposed \emph{sMARE}. Correlation coefficients remain one
of the most widely used metrics; they measure
the % strength and direction of 
association between the predicted and the
actual performance values, mostly in terms of Average
Precision~\cite{townsend_2002_clarity,cummins_qpp_sig-11,roitman_ictir17,kurland_tois12,uef_kurland_sigir10,zhou_2007_qpp}
across a set of queries. 
As discussed in Section~\ref{sec:intro}, Hauff et
al.~\cite{hauff2009combination} highlighted the drawbacks of using $\rho$, and
suggested reporting confidence intervals.

%% file: ecir25/results.tex
\section{Experimental Setup} 
\label{sec:exp}
% In this Section, we first state the specific research questions in relation to combining multiple performance predictors, following which we explain the datasets used and QPP methods investigated for this study. 

% \subsection{Research questions} \label{ss:rq}
% \textsc{Hauff et al.}~\cite{hauff2009combination} observe that combining different QP predictors, specifically pre-retrieval ones, can achieve significantly improved predictions in relation to rank correlation measures across different datasets. 
% However, they claim that the combination of predictors is not impactful in terms of improving RMSE. 
% In this paper, we revisit the study of~\cite{hauff2009combination} and extend the scope of our research with more recent datasets and a wide range of post-retrieval QPP approaches for a comprehensive study. Thus, we formulate our research questions as follows.

\input{ecir25/tabdefs/dataset}

\para{Datasets} 
For our investigation in this paper, we leverage three benchmark IR collections, TREC Robust (news articles with 250 topics), ClueWeb09B~\cite{clueweb} (crawled web pages with 200 topics), and MS MARCO Passage~\cite{msmarco-data} (a question answering dataset with over $100k$ Bing queries with 97 topics). 
% \sd{97 queries ta lekha thik hobe na bdhy, karon MS MARCO er dev set ei more than 6k queries}. 
Table~\ref{tab:datastats} gives summary statistics for these collections.
% , with the experimental topics highlighted in bold. 
% The topics based on which the experiments are performed in this study are highlighted in Table~\ref{tab:datastats} with boldface.
For the experiments on ClueWeb09B, 
spam documents are removed using the Waterloo spam filter~\cite{clueweb} with spam confidence $> 70\%$. %, and we name the subset as CW09B. 
TREC 678 with 150 queries is used to replicate the results reported by Hauff et al. (see Table \ref{tab:pre-ret-reprod-table}).
% Additionally, we employ newer collections such as ClueWeb09B and MS MARCO passage, which have gained popularity among researchers in QPP studies.
Additionally, our preliminary analysis included two other web collections -WT10G~\cite{hawking2000trecweb} and {Gov2}~\cite{clarke:trecterabyte2004} (also used in~\cite{hauff2009combination}) - which yielded trends consistent with those found in the newer web collections in our study.
% In our preliminary analysis, we also experimented with the two additional web datasets - WT10G~\cite{hawking2000trecweb} and Gov2~\cite{clarke:trecterabyte2004}, and observed consistent trends with the newer and robust collections used in our study.
% Due to space limitations in this paper, we have excluded those initial findings and have only reported results on the TREC Robust alongside more recent and widely used web collections, such as ClueWeb09B and MS MARCO passage, which have gained popularity among researchers in QPP studies.
% 
Due to space limitations, we exclude those initial findings in this paper and instead, focus on the TREC Robust alongside more recent and widely used web collections, ClueWeb09B and MS MARCO passage, which have gained popularity among researchers in QPP studies.

\begin{comment}
    
It is worth mentioning here that \textsc{Hauff et al.} used one news collection (TREC 678) and two web collections (WT10G~\cite{hawking2000trecweb} and Gov2~\cite{clarke:trecterabyte2004}) for their experiments.
In our study, we replicate their experiments on the TREC 678 collection (see Table \ref{tab:pre-ret-reprod-table}). We also report the extended performance on the Robust collection with 100 more queries in Table~\ref{tab:regr-pre-ret-table}. 
We further extend the evaluation to ClueWeb09B, a considerably larger Web collection, as well as the more modern MS MARCO passage dataset which is more suitable for a wide range of recently proposed neural QPP approaches.
% , such as \texttt{BERT-QPP}~\cite{bertqpp}, \texttt{qppBERT-PL}~\cite{qppbertpl_datta22}, \texttt{BERTPE}~\cite{bertpe} etc.
% 
Note that, for our experimental setup, we combine all queries from the TREC Robust and TREC DL'19, '20 collections (see Table \ref{tab:datastats}). In subsequent tables, we refer to these as TREC Robust and TREC DL, respectively. 
For the experiments with ClueWeb09B, spam documents are removed using the Waterloo spam scores~\cite{clueweb} with spam confidence $> 70\%$, and we name the subset as CW09B. 
\end{comment}
 
\para{Pre-retrieval QPP methods} 
% In this study, we have selected 10 of the most effective pre-retrieval predictors, identified based on the empirical analysis presented in~\cite{hauff2009combination}.
% Specifically, 
We employ a range of pre-retrieval predictors targeting distinct aspects of query performance. 
Firstly, the specificity-based methods that include \texttt{AvgIDF}, \texttt{MaxIDF}, \texttt{SumSCQ}, \texttt{AvgSCQ}, and \texttt{MaxSCQ} which estimate query difficulty based on the informativeness of terms in the query.
Complementing these methods, we employ the 
ambiguity-based techniques, such as, \texttt{AvP} and \texttt{AvNP},
designed to capture the potential variability or ambiguity inherent in query terms.
Additionally, we incorporate rank sensitivity-based methods, like \texttt{SumVAR}, \texttt{AvgVAR} and \texttt{MaxVAR}~\cite{qpp_preret_ecir08} (see Section~\ref{ss:qpp}), which assess the sensitivity of document rankings to individual terms in the query.
Note that, these choices are made based on the finding reported in~\cite{hauff2009combination} where the authors demonstrated the strong individual contribution in the fused predictors (refer to Figure 6 in~\cite{hauff2009combination}).

\para{Post-retrieval QPP methods} We select a representative set of 10 post-retrieval predictors, grouped into three categories based on the level of supervision. Of these, six QPP methods, such as \texttt{NQC}~\cite{kurland_tois12}, \texttt{Clarity}, \texttt{WIG}~\cite{zhou_2007_qpp}, and \texttt{UEF}~\cite{uef_kurland_sigir10} with the three base estimators (i.e., \texttt{NQC}, \texttt{WIG} and \texttt{Clarity}) are unsupervised. 
On the other hand, we employ weakly supervised 
neural QPP approach, namely 
\texttt{NeuralQPP}, 
that learns the relative importance of different estimators to obtain an optimal feature combination. 
Additionally, we choose three other 
fully supervised convolutional-based, \texttt{DEEP-QPP} 
that leverages information from the sematic interactions between the terms in the top documents and the query; 
and transformer-based state-of-the-art models, \texttt{BERT-QPP} which is a cross-encoder based pointwise QPP estimator; and \texttt{qppBERT-PL} which also makes use of cross-encoding based BERT vectors generated for the corresponding query and document terms (for more details see Section~\ref{ss:qpp}). 
% It is worth mentioning that both \texttt{BERT-QPP} and \texttt{qppBERT-PL} are two state-of-the-art neural QPP models. 

\para{Train-test splits} In the QPP literature (e.g.,~\cite{kurland_tois12,neuralqpp,query_variants_kurland,datta23_tois_wrig}), the most common experiment setup usually involves random partitioning of the query set into two halves, %parts, 
\emph{train}, \emph{test} and repeating the process $30$ times. %$N$ times.
% For each partition, 
% The \emph{train} set is utilized to 
% tune the hyper-parameters to 
% find the free-parameter values of a predictor that yield optimal prediction performance (as measured by Pearson's $\rho$ or Kendall's $\tau$)
% The optimal set of parameters is then applied to the corresponding \emph{test} set to evaluate the effectiveness. 
% This process is repeated $N$ times, and 
The average outcome over $30$ splits is reported.  
An identical setup is used for our experiments with TREC Robust and
CW09B collections. %with $N$ set to $30$. %, as there is no dedicated
% training data for these datasets.
%In general, we will refer to this technique as a \emph{cross
%  validation} (\emph{CV}) approach.
However, because the MS MARCO collection has a designated train:test partition, we tuned the model hyper-parameters on a random sample (specifically $10\%$ of train split) and reported results on the TREC DL, which is the subset of MS MARCO test set, as prescribed by~\cite{bertqpp,qppbertpl_datta22,datta23_tois_wrig}. 
All the parameters are tuned by dividing the training set into $k$-fold cross-validations. 
Additionally, to align our results with those of Hauff et al., we
employ the same \emph{leave-one-out} sampling strategy to predict the
AP measure (see Table~\ref{tab:pre-ret-reprod-table}).

\para{Penalized regression} 
Our objective is to combine $m$ predictors $\{X_{1}, \ldots, X_{m}\}$, to predict the target variable, $Y$, which represents the average precision. 
A simple approach is to assume that the predictors are independent and apply multiple linear regression via Ordinary Least Squares (OLS). 
OLS aims to minimize the sum of squared residuals between the observed values and the predicted values, providing an estimate of the relationship between the predictors and the target variable.
However, in practice, it is often the case that the predictors are not independent, and may be correlated with each other. 
When such correlations exist, the OLS method may suffer from issues such as multicollinearity, leading to unstable and overfit models.
To address this issue, regularization techniques are commonly employed to improve the generalization ability of the models and enhance interpretability by enforcing sparsity.
The two most common forms of regularization are the $l_1$ and $l_2$ penalties. 
The $l_1$ penalty, known as LASSO (Least Absolute Shrinkage and Selection Operator), encourages sparsity by shrinking some of the coefficients exactly to zero, thus effectively selecting a subset of the most important predictors.
On the other hand, the $l_2$ penalty, associated with Ridge regression, encourages smaller coefficients overall but does not set any of them to zero, allowing all predictors to contribute to the model.
An even more flexible approach is the ElasticNet (denoted as \texttt{E-Net}), which combines both the $l_1$ and $l_2$ penalties, offering a balance between the sparsity-inducing properties of LASSO and the stability advantages of Ridge regression. The ElasticNet is particularly useful when there are many correlated predictors, as it can handle both the selection of predictors and the regularization of those predictors in a way that neither LASSO nor Ridge can achieve individually.
% If features are correlated, $l_1$ and/or $l_2$ regularizers may be used to improve generalization and sparsity. LASSO, Ridge, and ElasticNet (\texttt{E-Net}) are examples of these ``penalized regression'' approaches.
% OLS minimizes the squared distance between predicted AP and the actual AP measure.

Following~\cite{hauff2009combination}, we also use \texttt{BOLASSO}~\cite{bolasso}, a computationally inexpensive and
generalized version of \texttt{LASSO} that uses bootstrap samples, and \texttt{LARs}~\cite{efron_2004}.
% of LASSO (\texttt{BOLASSO}). Additionally, a computationally inexpensive and generalized version of LASSO, LARs~\cite{efron_2004} was 
% also investigated. 
The regularization parameters of LARs may be estimated either by introducing random predictors into the model (\texttt{LARS-Traps}) or through cross-validation (\texttt{LARS-CV}). 
The same set of regression methods were also employed in~\cite{hauff2009combination}.
% \textcolor{red}{finalize the set of regressors and change accordingly}. \textcolor{red}{TODO: decide Ridge ta jaba ki jabena.. LARs er ek -adh line}
% \footnote{Source code (including the QPP methods) and the data (runs and prediction scores) are available here : https://anonymous.4open.science/r/qpp-comb-2F03/}
%In order to combine different QPP methods, 
% Merged to Train-test split section
%\para{Sampling strategies}: \textcolor{red}{SS: check this. train:test splits section is added earlier. should we combine these two?} %We use the standard sampling strategies recently adopted by QPP researchers, in which the query set is randomly split into a 50:50 fold and the experiment is repeated 30 times. 
%All the parameters are tuned on by dividing the training set into $k$-fold cross-validations. Additionally, to compare the results with Hauff et al., we also apply a leave-one-out based sampling strategy to predict the AP measure. 

\para{Retrieval settings and evaluation} To ensure a fair comparison with Hauff et al., we use a language model with Dirichlet smoothing~\cite{lmdir} ($\mu$ set to $1000$) to retrieve the documents and measure AP with the top $1000$ documents. 
Note that, this is also the most popular choice among researchers working on the QPP domain~\cite{townsend_2002_clarity,cummins_qpp_sig-11,roitman_ictir17,kurland_tois12,uef_kurland_sigir10,zhou_2007_qpp}.
As the ranges of QPP scores vary across different predictors, we normalize them with min-max normalization before applying penalized regression. 
% \inote{\textbf{TODO: }\red{mention that in our study, we use rho, tau, etc. measures.}}
%For QPP evaluation, we employ the commonly used correlation measures - Pearson's $\rho$  and Kendall's $\tau$ (henceforth denoted as $\rho$ and $\tau$, respectively). %While the former is a value-based correlation, the latter is a rank-based one. 
%We also evaluate our results with \emph{RMSE} as prescribed by Hauff et al. and \emph{sMARE}, a rank error based measure as described in Section~\ref{ss:eval}.  
For evaluation, we employ the commonly used correlation measures - Pearson's $\rho$, Kendall's $\tau$ ($\rho$ and $\tau$ respectively), \emph{RMSE}, and \emph{sMARE}. 

%% file: ecir25/tabdefs/dataset.tex
\begin{table}[t]
\centering
\caption{\small Datasets used in our experiments. `avg.$|Q|$' and `avg.\#rel' denote average query length and average number of relevant documents, respectively.
% The 
`Query ids' column for TREC DL is left empty as the topic
identifiers do not follow a particular order.
%The `Identifier' represents the topics on which the results are reported in subsequent tables.
}

\begin{adjustbox}{width=1\columnwidth}
\small
\begin{tabular}{@{}l@{~~~}l@{~~~}l@{~~~~}l@{~~~~}c@{~~}c@{~~}c@{~~}c@{}}
\toprule
Collection & Type & Document Set (\#docs) & Topic Set & Query ids & \#topics & avg.$|Q|$ & avg.\#rel \\\midrule

\textbf{TREC 678}    & \multirow{2}{*}{News} & Disks 4,5 minus CR & TREC 6, 7, 8 Adhoc topics & 301-450          & 150 & 2.45 & 89.13 \\ 
\textbf{TREC Robust} &                       & (528,155)          & TREC 678 + Robust topics  & 301-450, 601-700 & 250 & 2.62 & 68.36 \\\midrule

\multirow{2}{*}{\textbf{CW09B}} & \multirow{2}{*}{Web} & ClueWeb09B-S70 & \multirow{2}{*}{TREC Web Track 2009--2012 topics} & \multirow{2}{*}{1-200} & \multirow{2}{*}{200} & \multirow{2}{*}{2.42} & \multirow{2}{*}{16.02} \\
& & (29,038,220) & & & & \\\midrule

\multirow{2}{*}{\textbf{TREC DL}} & \multirow{2}{*}{Web} & MS MARCO Passage & \multirow{2}{*}{TREC DL'19 + '20 topics} & \multirow{2}{*}{--} & 97 & \multirow{2}{*}{5.76} & \multirow{2}{*}{42.96} \\
& & (8,841,823) & & & (43+54) & \\

\bottomrule
\end{tabular}

\label{tab:datastats}
\end{adjustbox}
\end{table}

%% file: ecir25/analysis.tex
\section{Experimental Results} \label{sec:analysis}

\input{ecir25/tabdefs/pre-ret}
\subsection{Reproducing results from \cite{hauff2009combination}}
As a validation check, we first repeat Hauff et al.'s experiments (with a
selected subset of the pre-retrieval predictors) on the TREC 678 topics
(see Table~\ref{tab:datastats}). Our results are presented in
Table~\ref{tab:pre-ret-reprod-table}. We note that our $\rho$ values are
generally higher than those reported by Hauff et al. Following their
suggestion, we include the 95\% confidence interval (CI) for the observed
$\rho$ values. For two predictors (\texttt{AvP}, \texttt{AvNP}), the CIs
(highlighted) are non-overlapping; since the corresponding $\rho$s are fairly
low on an absolute scale, we have not looked more carefully into this
difference. For \texttt{AvgSCQ} the difference (underlined) appears
numerically substantial, but is not, in fact, statistically significant. Of
greater concern is the noticeable difference in $\rho$ for \texttt{AvgIDF}
(highlighted): our measured $\rho$ is at one end of the CI reported in
\cite{hauff2009combination}, and vice versa. However, the RMSE values that
we obtain are all very close to those reported in
\cite{hauff2009combination}. We tentatively conclude that the
high-level summary of the results in Table~\ref{tab:pre-ret-reprod-table}
(that \texttt{MaxIDF}, \texttt{AvgIDF}, \texttt{AvgVAR} and \texttt{MaxVAR}
provide the best predictions, both in terms of $\rho$ and \emph{RMSE}) matches
the findings as in \cite{hauff2009combination}.

The lower part of Table~\ref{tab:pre-ret-reprod-table} shows results for
combined predictors. The values of $\rho$ reported in~\cite[Table
5]{hauff2009combination} for the combined approaches are, in some cases,
substantially (though not statistically significantly) higher than those
obtained via our implementation\footnote{\scriptsize{The combinations
  in~\cite{hauff2009combination} involved $19$ pre-retrieval predictors,
  while we combine $10$ of the best predictors from their set, but this
  does not seem to adequately explain these differences.}}. For Hauff
et al., combined predictors obtained upto $30\%$ higher $\rho$ than the best
performing singleton predictor; for us, these improvements are negligible
($< 2\%$). However, the improvements in terms of $\tau$ and \emph{sMARE} are a
little more noticeable.
In terms of \emph{RMSE}, the individual predictors perform roughly at par
with the combinations; this is consistent with Hauff et al. Overall,
Table~\ref{tab:pre-ret-reprod-table} indicates that, across evaluation
measures, combining predictors is not beneficial. This result is in
contrast to Hauff et al.'s observation that ``\emph{Independent of the
  quality of the predictors, $r$ [denoting Pearson's $\rho$] increases as more
  predictors are added to the model}''~\cite[Figure
2]{hauff2009combination}. Notably, their paper does not provide any further
insights into this surprising behaviour. In contrast, we offer an
explanation for this major difference in Section~\ref{sec:discussion}.

\input{ecir25/tabdefs/pre-combined}
\input{ecir25/tabdefs/post-combined}

\subsection{Pre-retrieval predictors on additional collections}
% \para{Pre-retrieval predictors on additional collections}
%
% \itodo{CIs, significance tests for Tables 2, 3}
% \red{Is it necessary to repeat the following here: In accordance with the
%   practice followed in the QPP research, we perform rest of the experiments
%   using the CV technique (see in Section~\ref{sec:exp}) on TREC Robust and
%   CW09B, and the designated train:test split on TREC DL collection.}
Table~\ref{tab:pre-ret-combined-table} presents the pre-retrieval QPP
results across three datasets in terms of all the evaluation measures.
% Kendall's $\tau$, Pearson's $\rho$, \emph{sMARE}, and \emph{RMSE}. 
The results for TREC DL are clearly in
agreement with the summary for Table~\ref{tab:pre-ret-reprod-table}: once
again, \texttt{MaxIDF}, \texttt{AvgIDF}, \texttt{MaxVAR}, and
\texttt{AvgVAR} appear to be the best-performing predictors in terms of all
metrics. This observation mostly holds for TREC Robust as well, where
\texttt{MaxSCQ} also emerges as a promising predictor, outperforming
\texttt{MaxIDF} and \texttt{AvgIDF} on two of the four metrics (underlined in Table~\ref{tab:pre-ret-combined-table}). CW09B presents the
most confusing picture: there is little variation in the \emph{RMSE} values
across all methods; to a lesser extent, the same holds true for
\emph{sMARE}. Overall, \texttt{MaxSCQ} seems to deliver the best results,
even though the \texttt{IDF} and \texttt{VAR} variants also perform
reasonably well.

As before, we present the penalized regression results in the lower part of
Table~\ref{tab:pre-ret-combined-table}. Once again, combining pre-retrieval
predictors does not improve results on any of the collections; in
fact, combined predictors often perform worse. For
example, for CW09B, Pearson's $\rho$ decreases from $0.2914$ (\texttt{MaxSCQ}, CI=[0.16,0.41])
to $0.1707$ (\texttt{BOLASSO}, CI=[0.03,0.30]), and \texttt{sMARE}, \texttt{RMSE}
scores increase marginally from $0.2660$ (\texttt{MaxIDF}) and $0.1625$
(\texttt{MaxSCQ}) to $0.2857$ (\texttt{OLS}) and $0.1661$ (\texttt{E-Net}),
respectively.
Thus, in response to \textbf{RQ1},
we find that combining predictors using penalized regression techniques \emph{does not} improve predictor performance (as measured by correlations with AP).
For more recent and larger collections, this could even lead to degraded performance.
For these newer collections, both \emph{RMSE} and \emph{sMARE} metrics also show a decline when compared to individual predictors.

% Table~\ref{tab:pre-ret-reprod-table} presents the pre-retrieval results for TREC 678 ($150$ queries) using a leave-one-out approach, where $\rho$ (Hauff et al.) and \emph{RMSE} (Hauff et al.) refer to the actual values of $\rho$ and \emph{RMSE} as reported in~\cite{hauff2009combination}. 
% To address \textbf{RQ-1}, the lower half of the Table~\ref{tab:pre-ret-reprod-table} reports
% %we report (second half of Table~\ref{tab:pre-ret-reprod-table})
% penalized regression values combining all pre-retrieval QP predictors as in Table~\ref{tab:pre-ret-combined-table}, where no substantial increase is observed in terms of either correlation or \emph{RMSE}, when comparing individual QPP measures with the combined approach.
%In the second row, we also report the penalized regression values combining all QP predictors. 
%Noticeably, when comparing individual QPP measures with the combined ones, there is no significant increase in correlation or \emph{RMSE}. 

% \para{Results for post-retrieval predictors.}
\subsection{Results for post-retrieval predictors}
Table~\ref{tab:post-ret-combined-table} shows the performance of individual
post-retrieval methods, as well as their combinations. We observe a clear
separation in performance between supervised and unsupervised methods,
with \texttt{qppBERT-PL} performing best overall.
% For example, on TREC Robust, among unsupervised approaches,
% \texttt{UEF-WIG} achieves the lowest \emph{RMSE} score ($0.1924$),
% while the \emph{RMSE} for \texttt{qppBERT-PL} is $0.1632$.
The bottom group in Table~\ref{tab:post-ret-combined-table} presents
results obtained by combining 10 post-retrieval predictors (these include
both unsupervised and supervised methods; please see Section~\ref{sec:exp}
for details). We observe that both \texttt{BOLASSO} and \texttt{E-Net}
approaches perform comparably well across all evaluation metrics. Compared
to the best standalone predictor (\texttt{qppBERT-PL}), these methods yield
notable improvements (23\% and 18\%, resp.) in $\tau$ and $\rho$ for the TREC
Robust collection, but only the decline in \emph{RMSE} (from $0.1632$
(\texttt{qppBERT-PL}) to $0.1371$ (\texttt{BOLASSO})) is statistically
significant (one-sided $t$-test, \emph{p}-value = $0.0328$). Further, these
improvements fade as collection size increases, dropping to less than 5\%
for larger collections. None of the improvements observed on CW09B and TREC
DL are significant. Indeed, in one case (rank correlation ($\tau$) on CW09B),
a performance degradation is observed. These findings address
\textbf{RQ2}, and suggest that pre- and post-retrieval predictors behave
differently. Specifically, combining post-retrieval predictors improves
prediction accuracy, although this advantage gradually declines as
collection size grows.

%% file: ecir25/tabdefs/pre-ret.tex
\begin{table}[t]
  \centering
  \caption{\small Comparison of our results with Hauff et al.'s, for pre-retrieval predictors on TREC 678. For reporting \emph{RMSE} and \emph{sMARE}, we use the leave-one-out based approach described in \cite{hauff2009combination}. Note that, for the regression methods,
    % for OLS and Lars-CV, 
    $\rho$ and \emph{RMSE} values are not directly comparable to those in
    \cite{hauff2009combination}, as they used 19 predictors while we use 10
    selected predictors. The `\% imp.' columns show the relative
    improvement in $\rho$ achieved by the regression methods over the best
    performing singleton predictor (\texttt{AvgIDF} for us, and
    \texttt{MaxIDF} in \cite{hauff2009combination}). Values of $\rho$ are
    highlighted or underlined if the value obtained by us is noticeably
    different from that in \cite{hauff2009combination}.}
  \label{tab:pre-ret-reprod-table}

% \begin{adjustbox}{width=.6\textwidth}
  \resizebox{1\columnwidth}{!}{
    \begin{tabular}{l | @{~}c@{~~}c@{~~}c@{~~}l@{~~}c@{~~}c@{~~}c@{~~}c} 
    \toprule
     % & \multicolumn{6}{c}{TREC 678}  \\
     % \cmidrule{2-7}
 QPP      & $\tau$ & $\rho$ $(CI)$ & \% imp. & $\rho$ ({Hauff et al.}) & \% imp. & \emph{sMARE} & \emph{RMSE} & \emph{RMSE} ({Hauff et al.}) \\
 
 \toprule
 
 \texttt{MaxIDF}   & 0.4085 & 0.5780  [0.46,0.68] & - & 0.53 [0.41,0.64]           &  -  & 0.1937       & 0.1766      & 0.186                               \\
 
 \texttt{AvgIDF}   & 0.3897 &  \colorbox{green!15}{0.6283}  [0.52,0.72] & - & \colorbox{green!15}{0.52} [0.39,0.62]    &  -  & 0.1997       & 0.1674      & 0.188                               \\
 
 \texttt{SumSCQ}   & 0.0525 & -0.0162 [-0.18,0.14] & - & 0.00 [-0.16,0.16]     & -  & 0.3123       & 0.2155      & 0.217                               \\
 
 \texttt{MaxSCQ}   & 0.3825 & 0.4093  [0.27,0.53] & - & 0.34 [0.19,0.47]       &   -      & 0.2096       & 0.1953      & 0.205                               \\
 
 \texttt{AvgSCQ}   & 0.2930 & \ul{0.3691}  [0.22,0.50] & - & \ul{0.26} [0.10,0.40]     &   -        & 0.2252       & 0.1999      & 0.210                               \\
 
 \texttt{SumVAR}   & 0.2855 & 0.3186  [0.17,0.46] & - & 0.30 [0.14,0.44]      &    -      & 0.2383       & 0.2060      & 0.206                               \\
 
 \texttt{AvgVAR}   & 0.4327 & 0.5933  [0.48,0.69] & - & 0.51 [0.38,0.62]      &   -       & 0.1891       & 0.1745      & 0.185                               \\
 
 \texttt{MaxVAR}   & 0.4505 & 0.5840  [0.47,0.68] & - & 0.51 [0.38,0.62]      &    -      & 0.1840       & 0.1780      & 0.182                               \\
 
 \texttt{AvP}      & 0.2122 & \colorbox{green!15}{0.3197}  [0.17,0.46] & - & \colorbox{green!15}{-0.12} [-0.28,0.04]     &    -    & 0.2732       & 0.2039      & 0.214                               \\
 
 \texttt{AvNP}     & 0.1490 & \colorbox{green!15}{0.1965}  [0.04,0.35] & - & \colorbox{green!15}{-0.22} [-0.37,-0.06]    &   -       & 0.2909       & 0.2109      & 0.210                               \\
 
 \bottomrule
 
 \texttt{OLS}      & 0.4619 & 0.6205 [0.51,0.71] & -1.24 & {0.69} [0.60, 0.77]   &  30.19   & 0.1905       & 0.1690      & {0.188}            \\
 
 \texttt{LARS-Traps} & 0.4297 & 0.6328 [0.53,0.72] & 0.72 & {0.59} [0.47, 0.68]  &  11.32    & 0.1953 & 0.1702 & {0.179} \\
 
 \texttt{LARS-CV}  & 0.4485 & 0.6398 [0.53,0.73]  & 1.83 &  {0.68} [0.59, 0.76]  &  28.30    & 0.1897       & 0.1642      & {0.183}           \\
% LASSO-CV & 0.4487 & 0.6357  &  -                            & 0.1955       & 0.1641      & -                                \\
% Ridge-CV & 0.4385 & 0.5968  &  -                             & 0.1972       & 0.1708      & -                                \\
\texttt{BOLASSO} & 0.4504 & 0.6365 [0.53,0.72] & 1.30 & {0.59} [0.47, 0.68]       &   11.32 & 0.1917      &  0.1641      & {0.181} \\
% BOLASSO: number of bootsrap examples : 10 (as mentioned in Hauff et al.), earlier reported figure was with the 128 samples. 
 \texttt{E-Net} & 0.4444 & 0.6253 [0.52,0.71] & -0.48 &  {0.69} [0.60, 0.77]       &  30.19  & 0.1966       & 0.1660      & {0.182}          \\
 \bottomrule

    \end{tabular}
    }
  % \end{adjustbox}
\end{table}

%% file: ecir25/tabdefs/pre-combined.tex
\begin{table}[t]
  \centering
    \caption{\small Results for pre-retrieval methods (top group), and
      their combinations (bottom group) across three datasets. 
      We report \emph{p}-values for a one-tailed $t$-test when comparing \emph{RMSE}s of the combined predictors with the best-performing individual QPP method, i.e., \texttt{MaxIDF}. The best values for each
    metric--collection pair are in bold face. 
    % Pre-retrieval results across three datasets. 
    % \textcolor{green}{Final}
  }
  \label{tab:pre-ret-combined-table}

\begin{adjustbox}{width=1\textwidth}
\begin{tabular}{l |c c c c c | c c c c c | c c c c c} 
\toprule
% \multicolumn{13}{c}{Pre-Retrieval}\\\toprule
 & \multicolumn{5}{c}{TREC Robust} 
 & \multicolumn{5}{c}{CW09B} 
 & \multicolumn{5}{c}{TREC DL} \\
 
 \cline{2-16}
 
Predictor & $\tau$ & $\rho$ & \emph{sMARE} & \emph{RMSE} & \emph{p-value} & $\tau$ & $\rho$ & \emph{sMARE} & \emph{RMSE} & \emph{p-value} & $\tau$ & $\rho$ & \emph{sMARE} & \emph{RMSE} & \emph{p-value} \\\toprule

\texttt{MaxIDF} & 0.3269 & 0.4020 & 0.2264 & 0.1946 & - & 0.2017 & 0.2239 & \textbf{0.2660} & 0.1665 & -  & \textbf{0.3267} & \textbf{0.5053} & 0.2285 & \textbf{0.2270} & -     \\

\texttt{AvgIDF} & 0.2959 & \textbf{0.4702} & 0.2383 & \textbf{0.1867} & -  & 0.1506 & 0.1569 & 0.2853 & 0.1684 & -  & 0.3204 & 0.4884 & \textbf{0.2236} & 0.2276 & -     \\

%AvQC \\
%AVQCG  \\

\texttt{SumSCQ} & 0.0977 & 0.0617 & 0.3010 & 0.2116 & -  & 0.1635 & 0.2011 & 0.2839 & 0.1678 & -  & -0.0498 & -0.1573 & 0.3535 & 0.2580 & -     \\

\texttt{MaxSCQ} & \ul{0.3464} & 0.3787 & \ul{0.2194} & 0.1955 & - & \textbf{0.2142} & \textbf{0.2914} & 0.2670 & \textbf{0.1625} & -  & -0.0468 & -0.1733 & 0.3552 & 0.2622 & -    \\

\texttt{AvgSCQ} & 0.2550 & 0.3409 & 0.2539 & 0.1983 & - & 0.1472 & 0.1853 & 0.2896 & 0.1666 & - & 0.1198 & 0.0719   & 0.2921 & 0.2695 & -    \\

\texttt{SumVAR} & 0.2852 & 0.2810 & 0.2450 & 0.2070 & - & 0.1869 & 0.2426 & 0.2768 & 0.1666 & - & 0.0825 & 0.0853   & 0.3178 & 0.2606  & -   \\

\texttt{AvgVAR} & \textbf{0.3726} & 0.4647 & \textbf{0.2146} & 0.1887 & - & 0.1595 & 0.2002 & 0.2874 & 0.1666 & - & 0.3071 & 0.4093  & 0.2328 & 0.2387 & -    \\

\texttt{MaxVAR} & 0.3694 & 0.4575 & 0.2148 & 0.1913 & - & 0.1923 & 0.2761 & 0.2792 & 0.1639 & - & 0.2728 & 0.4069   & 0.2572 & 0.2395 & -    \\

\texttt{AvP}    & 0.1426 & 0.2432 & 0.2885 & 0.2052 & - & 0.0787 & 0.0611 & 0.3173 & 0.1695 & - & 0.1194 & 0.2072   & 0.2972 & 0.2558 & -    \\

\texttt{AvNP}   & 0.0802 & 0.1203 & 0.3034 & 0.2096 & - & 0.0262 & 0.0170 & 0.3197 & 0.1700 & - & 0.0755 & 0.1160   & 0.3014 & 0.2601 & -    \\

\midrule

\texttt{OLS} & \textbf{0.3389} & \textbf{0.4746} & \textbf{0.2250} & \textbf{0.1889} & 0.3606 & \textbf{0.1532} & 0.1304 & \textbf{0.2857} & 0.1789 & 0.7048 & 0.2009 & 0.3158 & 0.2694 & 0.2607 & 0.8959 \\

%\texttt{LASSO-CV} & 0.3209 & 0.4146 & 0.2309 & 0.1923 & 0.4376 & 0.1244 & 0.1471 & 0.2936 & 0.1662 & 0.4946 & 0.2471 & 0.3761 & 0.2533 & 0.2393 & 0.6871  \\

%\texttt{Ridge-CV} & 0.3383 & 0.4229 & 0.2253 & 0.1910 & 0.4040 & 0.1512 & \textbf{0.1763} & 0.2867 & \textbf{0.1657} & 0.4833 & 0.2405 & 0.3540 & 0.2561 & 0.2416 & 0.7324     \\
\texttt{LARS-Traps} &  0.3002 & 0.3737 & 0.2353 & 0.1957 & 0.5297 & 0.1319 & 0.1433 & 0.2932 & 0.1666 & 0.5027 & 0.2422 & 0.3605 & 0.2546 & 0.2394 & 0.7069 \\

\texttt{LARS-CV} & 0.2997 & 0.3851 & 0.2385 & 0.1959 & 0.5049 & 0.1141 & 0.1403 & 0.2955 & 0.1665 & 0.5002 & \textbf{0.2524} & \textbf{0.3876} & 0.2529 & \textbf{0.2387} & 0.6779 \\

\texttt{BOLASSO} & 0.3285 & 0.4367 & 0.2291 & 0.1913 & 0.4196 & 0.1500 & \textbf{0.1707} & 0.2872 & 0.1682 & 0.5359 & 0.2154 & 0.3311 & 0.2628 & 0.2530 & 0.8402 \\

\texttt{E-Net} & 0.3276 & 0.4226 & 0.2289 & 0.1912 & 0.4096 & 0.1307 & 0.1517 & 0.2923 & \textbf{0.1661} & 0.4921 & 0.2492 & 0.3766 & \textbf{0.2524} & 0.2391 & 0.6832  \\

\bottomrule
  \end{tabular}
  \end{adjustbox}
\end{table}

%% file: ecir25/tabdefs/post-combined.tex
\begin{table}[t]
  \centering
  \caption{\small Results for unsupervised (top
    group) and supervised (middle group) post-retrieval methods, as well as
    their combinations (bottom group), across three datasets. We
    report \emph{p}-values for a one-tailed $t$-test when comparing \emph{RMSE}s of the combined
    predictors with the best-performing individual predictor,
    \texttt{qppBERT-PL}. The best values are in bold face.
    % We consider LM-DIR-based retrieval, and the retrieval depth $k$ is
    %set to 100. % \textcolor{green}{final}}
  }
  \label{tab:post-ret-combined-table}
  
  \begin{adjustbox}{width=1\textwidth}
  
  \begin{tabular}{l |c c c c c|c c c c c|c c c c c} 
  \toprule
    \multirow{2}{*}{Predictor} & \multicolumn{5}{c}{TREC Robust} & \multicolumn{5}{c}{CW09B} & \multicolumn{5}{c}{TREC DL} \\
    \cline{2-16}
  & $\tau$   & $\rho$   & \emph{sMARE} & \emph{RMSE} & \emph{p-value} & $\tau$  & $\rho$   & \emph{sMARE} & \emph{RMSE}  & \emph{p-value} & $\tau$ & $\rho$ & \emph{sMARE} & \emph{RMSE} & \emph{p-value} \\
 \cline{1-16}

\texttt{NQC}         & 0.3960 & 0.3315 & 0.2025 & 0.2025 & -  & 0.2423 & 0.1783 & 0.2554 & 0.1726 & -  & 0.3119 & 0.2463 & 0.2330 & 0.2577 & -   \\

\texttt{WIG}         & 0.3163 & 0.3982 & 0.2367 & 0.1950 & -  & 0.2624 & 0.3032 & \textbf{0.2447} & 0.1620 & - & 0.3295 & 0.4768 & 0.2315 & 0.2304 & -   \\

\texttt{Clarity}     & 0.2632 & 0.3899 & 0.2515 & 0.1948 & -  & 0.2244 & 0.3134 & 0.2693 & 0.1610 & - & 0.2796 & 0.4393 & 0.2459 & 0.2358 & -   \\

\texttt{UEF-NQC}     & 0.3801 & 0.3469 & 0.2066 & 0.2015 & -  & 0.2489 & 0.1853 & 0.2541 & 0.1711 & - & 0.3316 & 0.3092 & 0.2272 & 0.2505 & -  \\

\texttt{UEF-WIG}     & 0.2938 & 0.4205 & 0.2430 & 0.1924 & -  & \textbf{0.2697} & 0.1069 & 0.2467 & 0.1789 & -  & 0.3402 & 0.4887 & 0.2279 & 0.2299 & -   \\

\texttt{UEF-Clarity} & 0.2313 & 0.2962 & 0.2614 & 0.2019 & -  & 0.1611 & 0.2881 & 0.2887 & 0.1664 & -  & 0.2822 & 0.4560 & 0.2470 & 0.2325 & -   \\

\cmidrule{1-16} 

\texttt{NeuralQPP}   & 0.4204 & 0.3165 & 0.2031 & 0.2003 & -  & 0.1995 & 0.3054 & 0.2691 & 0.1619 & - & 0.3857 & 0.5287 & 0.2098 & 0.2212 & -   \\

\texttt{qppBERT-PL}  & \textbf{0.4740} & \textbf{0.6396} & \textbf{0.1759} & \textbf{0.1632} & -  & 0.2284 & \textbf{0.3553} & 0.2567 & \textbf{0.1591} & -  & 0.3905 & \textbf{0.5939} & 0.2030 & \textbf{0.2113} & -   \\

\texttt{Deep-QPP}    & 0.4348 & 0.5598 & 0.1874 & 0.1747 & -  & 0.2201 & 0.3332 & 0.2686 & 0.1604 & - & 0.3892 & 0.5492 & 0.2151 & 0.2176 & -  \\

\texttt{BERT-QPP}    & 0.4656 & 0.6093 & 0.1808 & 0.1680 & -  & 0.2233 & 0.3424 & 0.2699 & 0.1598 & - & \textbf{0.4111} & 0.5459 & \textbf{0.1992} & 0.2205 & -  \\

\cmidrule{1-16}

\texttt{OLS}      & 0.5704 & 0.7246 & 0.1470 & 0.1469 & 0.1724 & 0.2325 & 0.3254 & 0.2610 & 0.1668 & 0.6494 & 0.4372 & 0.5946 & 0.1940 & 0.2146 & 0.5196   \\

\texttt{LARS-Traps} & 0.5639 & 0.7240 & 0.1507 & 0.1454 & 0.1094 & 0.2176 & 0.3527 & 0.2730 & 0.1567 & 0.4435 & 0.4169 & 0.5753 & 0.2040 & 0.2115 & 0.4981  \\    

% \texttt{LASSO-CV} & \textbf{0.5849} & \textbf{0.7539} & \textbf{0.1439} & \textbf{0.1378} & 0.0360 & 0.2252 & 0.3474 & 0.2669 & 0.1577 & 0.4630 & 0.4458 & 0.6183 & 0.1924 & 0.2017 & 0.3310  \\

%\texttt{Ridge-CV} & 0.5814 & 0.7494 & 0.1442 & 0.1396 & 0.0531 & \textbf{0.2625} & \textbf{0.4096} & \textbf{0.2582} & \textbf{0.1534} & 0.3653 & \textbf{0.4721} & \textbf{0.6553} & \textbf{0.1837} & \textbf{0.1933} & 0.2025 \\

\texttt{LARS-CV} & 0.5595 & 0.7225 & 0.1512 & 0.1489 & 0.1524 & 0.1990 & 0.3073 & 0.2746 & 0.1720 & 0.5208 & 0.4471 & 0.6216 & 0.1917 & 0.2003 & 0.3032  \\    

\texttt{BOLASSO} & \textbf{0.5876} & \textbf{0.7577} & 0.1490 & \textbf{0.1371} & 0.0328 & \textbf{0.2486} & 0.3581 & \textbf{0.2576} & 0.1615 & 0.5462 & 0.4440 & 0.6171 & 0.1930 & 0.2071 & 0.4156   \\    

\texttt{E-Net} & 0.5842 & 0.7538 & \textbf{0.1442} & 0.1379 & 0.0347 & 0.2355 & \textbf{0.3652} & 0.2646 & \textbf{0.1563} & 0.4316 & \textbf{0.4508} & \textbf{0.6266} & \textbf{0.1907} & \textbf{0.1997} & 0.2949       \\

\bottomrule
\end{tabular}
\end{adjustbox}  

\end{table}

%% file: ecir25/analysis2-mm.tex
\input{ecir25/figs/heatmap}
\section{Discussion}
\label{sec:discussion}
We now turn to a deeper analysis of the behavior of combined predictors, by
looking more closely at the relationships \emph{between} different
predictors. To the best of our knowledge, this issue has received limited
attention, and seems to have been carefully explored only
in~\cite{hauff_thesis}, where the author reported Kendall's $\tau$ among a set
of pre-retrieval predictors (see Tables 2.3 and 2.6
in~\cite{hauff_thesis}).

We compute pairwise $\rho$ values between the lists of QPP scores calculated
using different QPP methods\footnote{\scriptsize 
% Similar trends were observed with respect to $\tau$, but details are omitted due to space constraints. 
We also validated our results for the TREC678 collection against those reported in~\cite{hauff_thesis}.}.
% to identify the underlying reasons for performance variations
% This analysis aims to unravel how different predictors interact and contribute to overall predictive accuracy.
Figure~\ref{fig:heatmap} displays these values as a heatmap for each
collection for the two families of QPP approaches. 
% The figure reveals several surprising behaviors in the relationships between predictors.
Notably, we observe a high correlation ($\rho > 0.7$) among the \texttt{IDF}
variants, \texttt{VAR} variants, collection-based specificity methods, and
the polysemy-based approaches (\texttt{AvP} and \texttt{AvNP}) across all
datasets when using pre-retrieval methods. Overall, the correlation among
the pre-retrieval predictors are seen to be moderate to high with the
exception of \texttt{SumSCQ} and \texttt{SumVAR}, which do not correlate
well with the other predictors. However, they are mutually strongly
correlated, with $\rho$ reaching a peak of $0.94$ in the case of CW09B.

In contrast, post-retrieval predictors generally show \emph{lower}
correlation between methods.
Among these, only the \texttt{UEF} variants display moderate to high correlation with each other.
Neural methods, particularly \texttt{NeuralQPP}, \texttt{qppBERT-PL}, \texttt{Deep-QPP}, and \texttt{BERT-QPP}, demonstrate varied correlation behavior; for collections with fewer training queries (e.g., TREC Robust and CW09B), these methods show negligible correlation, while their correlations increase with the availability of more training data, as observed with the TREC DL collection ($\rho > 0.5$).

Further, we observe nearly zero to negative intra-predictor correlations, particularly among the pre-retrieval predictors in the TREC DL collection.
For example, in TREC DL collection, correlations between the pairs (\texttt{SumSCQ}, \texttt{AvgIDF}) or (\texttt{SumSCQ}, \texttt{AvgVAR}) are observed to be negative ($\rho <-0.3$).
In addition, the post-retrieval predictors exhibit a significant reliance on the amount of training data, which leads to low correlations in collections with fewer training samples.

% A similar pattern appears in CW09B, where \texttt{AvP} and \texttt{AvNP} are negatively correlated with \texttt{SumSCQ}. However, for TREC 678 queries, we observe no significant negative correlations among predictors, which we attribute as one reason why combining pre-retrieval QPP approaches does not enhance performance and may even reduce it.
% \inote{nicher puro ta rewrite korte hobe.}

By integrating the findings from the heatmap in Figure~\ref{fig:heatmap} with the fused predictors presented in Tables~\ref{tab:pre-ret-combined-table} and~\ref{tab:post-ret-combined-table}, we propose the following three hypotheses.

\begin{hyp}[H\ref{hyp:first}] \label{hyp:first}
If the correlation between predictors is moderate to high, combining them will not significantly improve performance.
\end{hyp}

Based on this hypothesis, we can analyze the results in the tables,
particularly for pre-retrieval in TREC Robust and post-retrieval in CW09B,
to understand the lack of improvement in the combined predictors. The
correlation between the predictors in these two cases (the first and fifth
heatmaps in Figure~\ref{fig:heatmap}) are moderate to strong; thus,
combining predictors for TREC Robust (in
Table~\ref{tab:pre-ret-combined-table}) and CW09B (in
Table~\ref{tab:post-ret-combined-table}) does not yield any improvements.
This lack of significant improvement can be attributed to the fact that no new information is being added, as the predictors being combined are quite similar to each other (as indicated by the high correlation).

\begin{hyp}[H\ref{hyp:second}] \label{hyp:second}
Low correlation among predictors may lead to an improvement in performance when they are combined.
\end{hyp}
% \inote{Robust-post ar DL-post}

In the fourth and sixth heatmaps in Figure~\ref{fig:heatmap}, we observe lower correlations among the pairs of post-retrieval predictors for the TREC Robust and TREC DL collections.
Our second hypothesis \textbf{H2} suggests that in such cases, combining predictors could enhance performance.
The performance improvements arise because the combined predictors are relatively unrelated, creating opportunities for complementary strengths to enhance overall accuracy.
This is supported by the results in Table~\ref{tab:post-ret-combined-table}, where we note improvements in these two collections.

\begin{hyp}[H\ref{hyp:third}] \label{hyp:third}
 If the correlation between predictors is negative, combining them may degrade performance.
\end{hyp}
% \inote{CW09B-pre ar DL-pre}

When the pre-retrieval predictors are applied to the TREC DL collection, we observe significant negative correlations between pairs of predictors as shown in Table \ref{tab:pre-ret-combined-table}.
Our third hypothesis, \textbf{H3}, posits that combining predictors 
% of these types, 
that conflict with one another may degrade overall performance.
This decline occurs because negatively correlated predictors provide contradictory signals, which may cancel each other out or introduce noise into the combined prediction.
This is empirically verified in Table~\ref{tab:pre-ret-combined-table} for the TREC DL dataset, where all metrics against the combined predictors indicate a decline in performance.

Thus, in response to \textbf{RQ3}, we conclude that a discernible relationship, whether positive or otherwise, among the predictors is evident due to which they exhibit similar performance producing the observed correlations.
This is supported by the heatmap and result tables, and is formalized by the proposed hypotheses.

% Interestingly, when comparing pre and post QP predictors, we find that \texttt{MaxIDF} and \texttt{AvgIDF}, despite their simplicity, perform similarly to unsupervised post-retrieval approaches in terms of \emph{RMSE} across the three datasets. In CW09B, \texttt{MaxIDF} even performs comparably to neural QPP approaches with respect to \emph{sMARE}.
%of \texttt{MaxIDF} is even comparable to neural QPP approaches in terms of \emph{sMARE}. 

%%% Local Variables:
%%% mode: latex
%%% TeX-master: "main"
%%% End:

%% file: ecir25/figs/heatmap.tex
\begin{figure}[t]
\centering

% OLD
% % pre
% \includegraphics[width=.332\textwidth]{figs/robust_pre-crop.pdf}
% \includegraphics[width=.286\textwidth]{figs/cw09_pre-crop.pdf}
% \includegraphics[width=.362\textwidth]{figs/dl_pre-crop.pdf}
% % post
% \includegraphics[width=.364\textwidth]{figs/robust_post-crop.pdf}
% \includegraphics[width=.272\textwidth]{figs/cw09_post-crop.pdf}
% \includegraphics[width=.346\textwidth]{figs/dl_post-crop.pdf}
% 
% pre
\includegraphics[width=.329\textwidth]{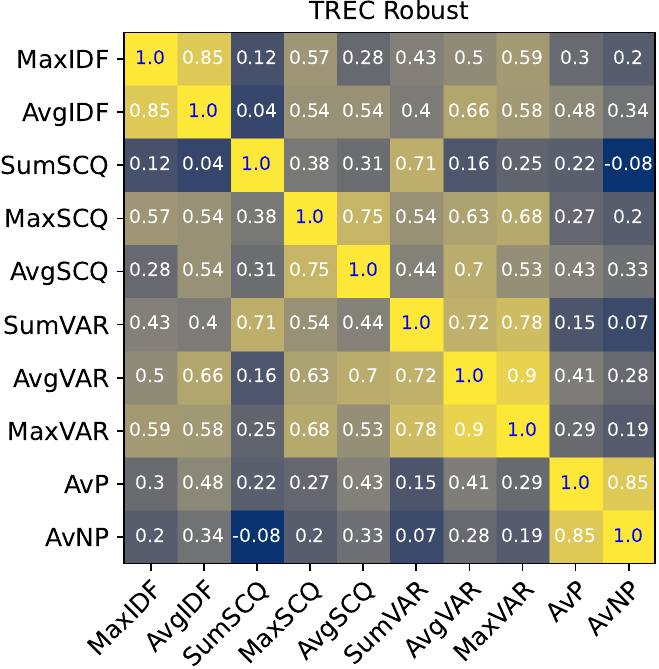}
\includegraphics[width=.2845\textwidth]{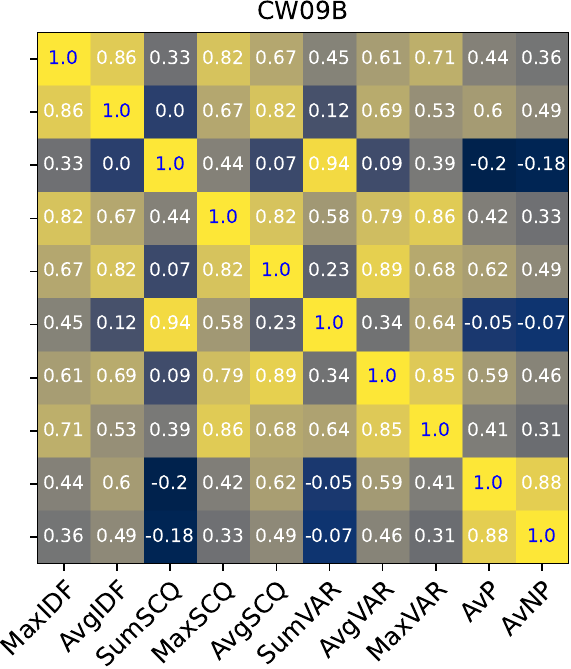}
\includegraphics[width=.365\textwidth]{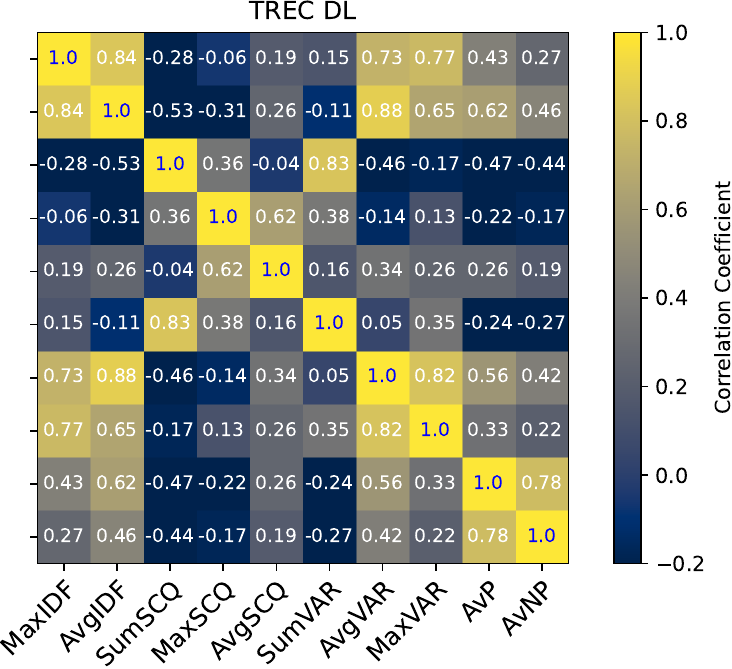}
% post
\includegraphics[width=.348\textwidth]{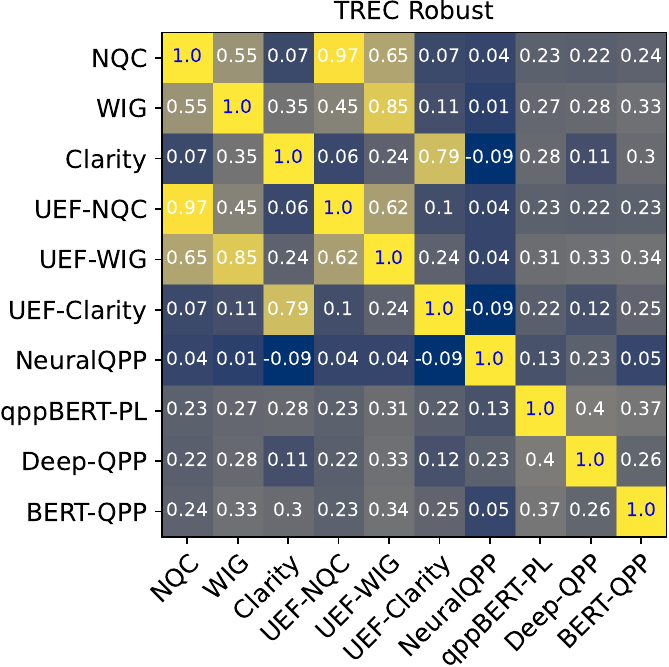}
\includegraphics[width=.271\textwidth]{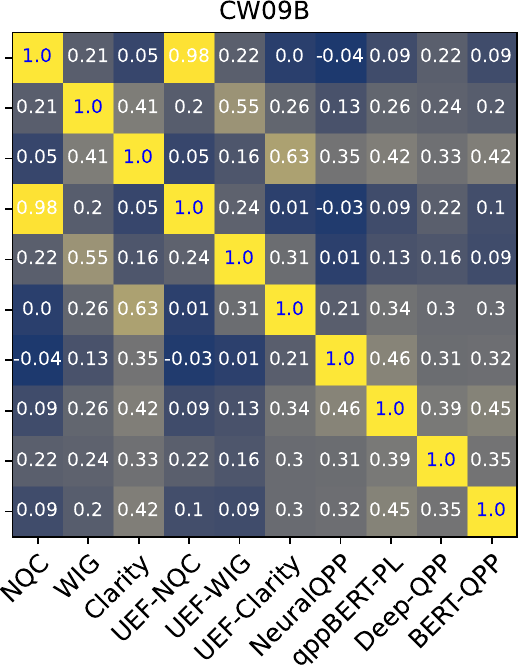}
\includegraphics[width=.357\textwidth]{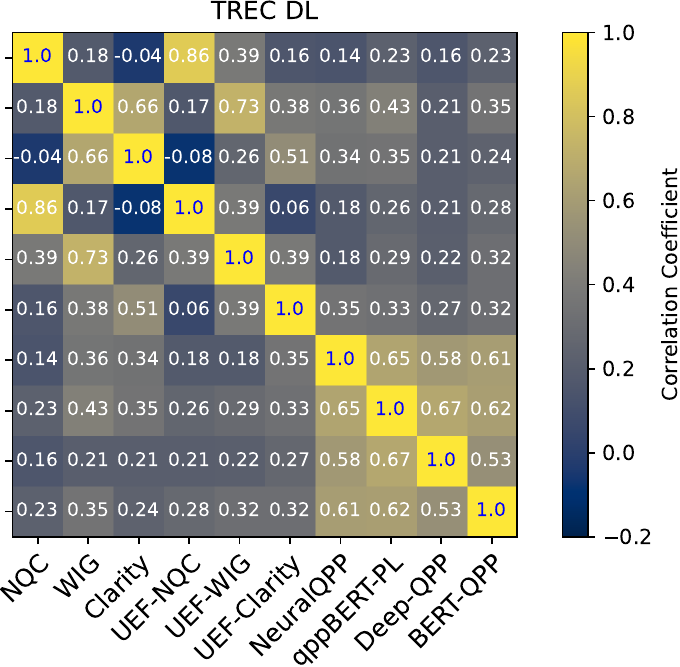}

\caption{
\small
% Heatmap visualizing the rank correlation among QPP methods based on their individual QPP scores. 
% Colour intensity represents the correlation values between the corresponding techniques measured with $\tau$ (lower triangle) and $\rho$ (upper triangle). The upper row depicts the correlation for pre-retrieval methods and the bottom row shows the post-retrieval QPP correlations for all the collections.
Heatmap visualizing the rank correlation among QPP methods based on their individual QPP scores. 
Colour intensity represents the correlation values between the corresponding techniques measured with $\rho$; lighter intensity represents higher correlation. The upper row depicts the correlation for pre-retrieval methods and the bottom row shows the post-retrieval QPP correlations for all the collections.
\label{fig:heatmap}
}

\end{figure}

%%% Local Variables:
%%% mode: latex
%%% TeX-master: "main"
%%% End:

%% file: ecir25/conclusion.tex
\section{Conclusion}
\label{conclusion}
In this study, we have revisited and extended earlier work on combining
different QPP methods to enhance prediction accuracy. Our results
demonstrate that, while combining predictors can improve correlation metrics
in some cases, these benefits diminish with larger collections and specific
predictor relationships. Notably, combining predictors that have low
positive correlation among themselves can yield improved performance, but
if the predictors are strongly positively correlated, or negatively
correlated, combining them provides no benefits, and may lead to degraded
results. This is a more fine-grained version of Hauff et al.'s observations.

Our experiments highlight that, while penalized regression techniques like
\texttt{BOLASSO} and \texttt{E-Net} may constitute robust combination strategies, the impact of combining predictors varies significantly between dataset types and predictor categories.
%This variance highlights the need for a nuanced approach to QPP, particularly as retrieval tasks grow in complexity and scale.
Thus, careful consideration of predictor relationships is essential to maximize QPP accuracy, with potential implications for future model selection and hybridization strategies in QPP research.
As part of future work, we plan to explore the individual combination process more carefully considering the predictor-predictor relationship.